\documentclass[sigconf]{acmart}

\AtBeginDocument{%
  }

\usepackage[utf8]{inputenc}
\usepackage{tabularx}
\usepackage{booktabs}
\usepackage{soul}
\usepackage{xcolor}
\usepackage{xspace}
\usepackage{graphicx}
\usepackage[caption=false]{subfig}  
\usepackage[export]{adjustbox}
\usepackage{amsmath}
\usepackage{textcomp}
\usepackage{listings}
\usepackage{cleveref}
\usepackage{algorithm}
\usepackage{balance}
\usepackage[noend]{algpseudocode}
\usepackage{enumitem}
\usepackage{multirow}
\usepackage{colortbl}
\usepackage{threeparttable}
\usepackage{multicol}
\usepackage{tcolorbox}

\definecolor{codegreen}{rgb}{0,0.6,0}
\definecolor{codegray}{rgb}{0.5,0.5,0.5}
\definecolor{codepurple}{rgb}{0.58,0,0.82}
\definecolor{backcolour}{rgb}{0.95,0.95,0.92}
\definecolor{gray}{gray}{0.9}
\definecolor{APA_stats}{RGB}{100, 100, 120}

\newcommand{\eg}{\emph{e.g.,}\xspace}

\newcounter{observation}

\copyrightyear{2024}
\acmYear{2024}
\setcopyright{acmlicensed}
\acmConference[IDE '24]{2024 First IDE Workshop}{April 20, 2024}{Lisbon, Portugal}
\acmBooktitle{2024 First IDE Workshop (IDE '24), April 20, 2024, Lisbon, Portugal}
\acmDOI{10.1145/3643796.3648446}
\acmISBN{979-8-4007-0580-9/24/04}

\begin{document}

\title{Context Composing for Full Line Code Completion}

\author{Anton Semenkin}
\email{anton.semenkin@jetbrains.com}
\affiliation{%
  \institution{JetBrains}
  \city{Belgrade}
  \country{Serbia}
}

\author{Yaroslav Sokolov}
\email{yaroslav.sokolov@jetbrains.com}
\affiliation{%
  \institution{JetBrains}
  \city{Berlin}
  \country{Germany}
}

\author{Evgeniia Vu}
\email{evgeniia.vu@jetbrains.com}
\affiliation{%
  \institution{JetBrains}
  \city{Bremen}
  \country{Germany}
}
\renewcommand{\shortauthors}{Semenkin et al.}

\begin{abstract}
    Code Completion is one of the most used Integrated Development Environment (IDE) features, which affects the everyday life of a software developer. 
    Modern code completion approaches moved from the composition of several static analysis-based contributors to pipelines that involve neural networks. 
    This change allows the proposal of longer code suggestions while maintaining the relatively short time spent on generation itself. 
    At JetBrains, we put a lot of effort into perfecting the code completion workflow so it can be both helpful and non-distracting for a programmer. 
    We managed to ship the Full Line Code Completion feature to PyCharm Pro IDE and proved its usefulness in A/B testing on hundreds of real Python users. 
    The paper describes our approach to context composing for the Transformer model that is a core of the feature’s implementation. 
    In addition to that, we share our next steps to improve the feature and emphasize the importance of several research aspects in the area.
\end{abstract}





\maketitle

\section{Introduction}
Code Completion is one of the most used IDE features~\cite{vs-usage-2016, eclipse-usage-2006}, which is present in the everyday life of a software developer. 
Modern code completion approaches moved from the composition of several static analysis-based contributors to pipelines that involve neural networks. 
The latter allows to propose longer code suggestions (\eg the whole line of code or even code snippet), while maintaining a relatively short time spent on generation itself. 
While multi-line suggestions are main-stream in AI-powered code completion, there is an evidence that single-line suggestions could be more preferable~\cite{cc-google, cc-meta} by developers. 
Our internal users' study also shows that multi-line completion is a situationally needed feature, however, single-line completion is handy often, which emphasizes its importance.

At JetBrains, we put a lot of effort into perfecting the code completion workflow, so it can be both helpful and non-distracting for a programmer.
In our Full Line Code Completion Team, the project scope is centered on generating code suggestions that are confined to a single line. 
To address potential privacy concerns associated with transmitting data to cloud providers, we have opted to work exclusively on the end user's local device. 
The model is run natively, so it enables us to adhere to strict execution time constraints, ensuring a smooth user experience. 
However, this local processing limitation imposes a cap on the size of the neural network model we can deploy. 
Our current estimations suggest that the upper bound is around 1 billion parameters, so latency and memory consumption remain reasonable. 
Consequently, maximizing the efficiency of the model within this size constraint is extremely important. 
Hence, context composition, also known as prompt engineering, becomes an essential aspect of the feature implementation.

Recently, our team has delivered the Full Line Code Completion (FLCC) feature to various IntelliJ-based IDEs, including PyCharm Pro~\cite{pycharm}.
The feature proved to be significantly useful in A/B testing and works seamlessly, naturally enhancing and improving the developer’s workflow. 
Our current approach enables integration into a developer's existing workflow without disrupting familiar patterns.

In this paper we are focusing on:
\begin{itemize}
    \item Description of the context composing setup, which proved to be useful for FLCC.
    \item Description of our upcoming experiments as well as methodologies that we did not find fruitful.
    \item Formulation of the open questions that require additional research in scope of neural code completion with models under 1B parameters.
\end{itemize}

\section{Full Line Code Completion Now}
\subsection{Current Setup}
In the core of the Full Line Code Completion feature, we use a GPT-like and LLaMA-like autoregressive language models, which predict a sequence of tokens based on the previous context. 
When code completion is invoked (either automatically or by user’s action), we compose the context for the model in the following manner.
\begin{itemize}
    \item \textbf{Pre-processing}. Source code from the current file is \textit{formatted}: leading whitespaces are removed and scope changes are replaced with special tokens; empty lines and trailing whitespaces are removed; comments are removed. 
    \item \textbf{Tokenization}. Obtained piece of code and a file path is tokenized with an approach based on byte-pair encoding (BPE), which is a common tokenization technique for Transformer-based neural networks.
    \item \textbf{Context construction}. We concatenate file extension, special token \texttt{<LANG\_SEP\_CHAR>}, file path, another special token \texttt{<METAINFO\_SEP\_CHAR>}, and the part of the code above caret, such that the total amount of tokens in the combined three parts fits the maximum context size of the model.
\end{itemize}

Since our scope is to work on the end user’s device, we care a lot about the efficiency of the composed context. 
Working on someone’s local machine imposes restrictions on working on a CPU with a relatively short context, so RAM usage of a used Transformer model does not drastically grow. 
So, among other implementation details that we came up with, we describe in more detail two important decisions.

Firstly, we do whitespaces trimming, which not only means empty lines removal but also deletion of leading whitespaces in each line.
For indentation-sensitive languages like Python leading whitespaces comprise an important part of the program because they indicate scope change.
So, for such languages we change each scope change with a corresponding special token: \texttt{<SCOPE\_IN>} and \texttt{<SCOPE\_OUT>}.
Thus, we avoid having tokens that differ only in the amount of whitespace in the beginning.
Typical examples of such tokens would be:
\begin{itemize}
    \item \texttt{for i in range(}
    \item \texttt{$\backslash$t for i in range(}
    \item \texttt{$\backslash$t $\backslash$t for i in range(}
\end{itemize}

Secondly, we do not use vanilla BPE for tokenization. 
Instead, we use ``long tokens'', which is basically a modified BPE approach that allows to merge tokens over space symbols but forbids merging over the end of the line. 
This approach allows us to create better representations of the lines in the source code and compress it more efficiently. 
For instance, the following recurring ``idioms'' in Python are compressed to a single token:
\begin{itemize}
    \item \texttt{for i in range(}
    \item \texttt{return True}
    \item \texttt{if \_\_name\_\_ == "\_\_main\_\_":}
\end{itemize}

Combining all of the details together, such an approach allows us to get the following information from the context: specific language dialect or file type it is being invoked in; general information about module and file specifics; and compact representation of the code above the caret to complete. 

With the described context composition we managed to create a code completion IDE feature that proved itself useful during A/B testing on hundreds of real users. 
Ratio of code completed (from total amount of code typed in the editor) for users with enabled Full Line Code Completion bumped 1.5 times compared to users that did not have the feature enabled. 
In addition to that, we have also received explicit feedback from tens of users. 
The feedback was mostly positive, reaching expressions that the feature is more preferable than some of the cloud-based large language model tools, but also described potential growth points. 
Users clearly want the feature to perform better given the existing context of their work, be that exact location in the current file, recently opened files, or the whole project. 
Our team managed to create a fast, useful and well-received feature that is already adopted by PyCharm Pro~\cite{pycharm} and DataSpell~\cite{dataspell}.

\subsection{Evaluation Setup}
To progressively enhance the Full Line Code Completion, we developed various feedback mechanisms, encompassing both offline and online evaluation methods. 
Offline evaluation is conducted using a dataset containing source code
The method often employed in academic settings due to the typical lack of access to a vast user base. 
Conversely, online evaluation is implemented by gathering data from actual users, making it a preferred approach for product development. 
In the FLCC, we initially assess the system's performance through offline evaluation; however, our emphasis predominantly lies on online evaluation, which we briefly describe.

At JetBrains, an early access program (EAP) is conducted several times per, allowing users to download ``experimental'' version of IDEs for free.
During EAPs, we split users to several groups, ship them different versions of some features and track a number of metrics for every group, also taking into an account statistical significance of the observed results.

Such an A/B testing was conducted by our team for code completion feature in PyCharm Pro.
During fall 2023, we not only studied users with and without FLCC, but also compared different implementations of FLCC itself.
Our golden star metric --- ratio of code completed (among all code written in the editor) bumbed 1.5 times for users with FLCC enabled compared to those with standard code completion only.
We also observed that users do not edit the selected code fragment immediately after inserting it into their code.
Finally, we did not observe any IDE performance degradation while enabling Full Line Code Completion.

\subsection{Our Planned Further Steps}
Besides releasing in other IntelliJ-based products, our next goal is to address the context composing issue to improve Full Line Code Completion even more. 

Recently, we started experimenting with bigger contexts than before, studying how the size of the prompt is connected with the quality of suggestions in our setup with relatively small neural networks. 
We have discovered that we can increase context several times without disturbing users too much with machine resource consumption. 
For PyCharm Pro release in December 2023 a model with a maximum context size of 384 tokens was used.
However, for the upcoming releases in 2024, we conducted experiments for models with a maximum context size of 1536 tokens. 
Such a change was possible because of the LLaMA usage instead of GPT.
According to our offline evaluation, we managed to increase code completion quality by 40\%, while keeping latency almost unchanged for most of real-world scenarios.
We continue conducting online evaluation of the implemented change, so we can assess its impact.

Longer contexts enable us to move forward to ``smart'' context composition techniques, which include fill-in-the-middle~\cite{incoder-paper} and retrieval-augmented generation~\cite{retro} methods. 
Given a bigger context size, one can include in the prompt not only lines above the actual caret position but also below it. 
Additionally, information can be retrieved from recently opened files, so the relevant code pieces are added to the context of the model.
We have already conducted some offline experiments with promising results (about 10\% target metric increase), so the next step is to polish relevant experience on the plugin side and conduct online experiments.

\section{Research Topics for Full Line Code Completion}
\subsection{Experiments We Conducted}
There is a ``small context'' issue known from users' feedback --- the model quickly ``forgets'' file content above the current typing position. 
This issue is not just limited to information at the beginning of a long file, but may also extend to methods of the current class that are significantly above the caret position. 
We came up with an approach that might mitigate this problem. 

The approach is based on the following idea. 
The most relevant context to the current typing position is the current method or function code combined with declarations of other methods and functions that exist in the scope. 
This approach reflects programmers' workflow: quickly get ideas about what’s implemented in the current class and pay attention to the closest code fragment. 
For implementation, we have composed the context in the following manner: 
\begin{itemize}
    \item Rearrange class methods or file functions so that they are placed before the current caret position
    \item Tokenize file, take current method or function tokens
    \item If context is not filled yet, concatenate methods declaration with already collected context
\end{itemize}

Despite following the intuition of a real person workflow, the approach did not show any positive results while experimenting. 
So, we abandoned this research direction.

\subsection{Open Problems}
We have a strong belief that filling context with relevant code is a fruitful point of growth for neural code completion systems (especially those working on the end user’s device). 
It is clear from large language models study that prompt engineering makes the difference, so we are looking forward to extensive research conducted in this field for not-so-large language models too. 
The area of most interest for us includes the following.
\begin{itemize}
    \item Study of the most beneficail code rearrangement techniques inside the same file.
    \item Research on most relevant files for context augmentation: last opened files, adjacent IDE tabs, adjacent project files, etc.
\end{itemize}

Programmers frequently switch between class implementation and class usage as well as they might develop several usages simultaneously, so the context of the recently used files matters. 
The key problem to solve the task is to collect relevant data to train models on. 
We surely cannot collect such data from IntelliJ users except for potential explicit opt-in in such a data collection program. 
Data can be potentially collected in anonymized form, training of such models can be performed in a federated manner or existing code completion models can be fine-tuned on a limited-size dataset collected by volunteers. 
We are looking for any ideas or implemented solutions in this direction.

Another promising direction for neural networks-based code completion systems is the RETRO-like methods~\cite{retro} that integrate a retrieval system into the Transformer model. 
Such systems allow automatic search of the relevant code pieces among fixed collection of source code files.
The obtained code pieces can then be used to enrich context of the Transformer model, so it generates better code suggestions. 
The extensive exploration of various methods for such a context enrichment represents a substantial interest for us.

We are aware that researchers often do not have access to datasets incorporating users' patterns of IDE usage.
For instance, combining datasets of adjacent IDE tabs presents a complex task, involving privacy concerns when the data is collected from real users.
We are open to collaborations that might result in the creation of datasets of a spoken nature, which could be useful for both researchers and practitioners while being fully legally compliant.
\section{Conclusion}
To sum up, our work on Full Line Code Completion at JetBrains highlights the effective use of a compact, context-aware GPT-like model, tailored for efficient operation on end-user devices. 
This approach has shown promising results in enhancing coding workflows within JetBrains' IDEs. 
Looking forward, we are excited about exploring more informative contexts, and we anticipate valuable contributions from the wider research community in advancing neural code completion systems.


\bibliographystyle{ACM-Reference-Format}
\bibliography{main_refs}


\end{document}